\documentstyle[prb,aps,preprint,psfig]{revtex}
\input{epsf.tex}
\include{psfig}
\begin{document}

\title{An Anderson-Fano Resonance and Shake-Up
  Processes in the Magneto-Photoluminescence of a Two-Dimensional
  Electron System}

\author{M. J. Manfra and  B. B. Goldberg}
\address{Dept. of Physics, Boston University, Boston MA 02215}

\author{L. Pfeiffer and K. West}
\address{Bell Laboratories, Lucent Technologies, Murray Hill, NJ 07974}

\date{\today}
\maketitle
\begin{abstract} 
  We report an anomalous doublet structure and low-energy satellite in
  the magneto-photoluminescence spectra of a two-dimensional electron
  system.  The doublet structure moves to higher energy with increasing
  magnetic field and is most prominent at odd filling factors $\nu=5$
  and $\nu=3$.  The lower-energy satellite peak tunes to {\it lower}
  energy for {\it increasing} magnetic fields between $\nu=6$ and
  $\nu=2$.  These features occur at energies below the fundamental band
  of recombination originating from the lowest Landau level and display
  striking magnetic field and temperature dependence that indicates a
  many-body origin.  Drawing on a recent theoretical description of
  Hawrylak and Potemski, we show that distinct mechanisms are
  responsible for each feature.
\end{abstract}
\newpage

Over the last decade a substantial body of work has focused on the
investigation of the photoluminescence spectra originating from the
recombination of a two-dimensional electron system with a valence-band
hole in the quantum Hall regime.  In conjunction with transport
measurements, much of this work has concentrated on examining the nature
of the incompressible ground states that arise at integral and
fractional filling.\cite{bbg-prl90} More recently, optical techniques
have proven sensitive to excited states of the 2DEG inaccessible to
transport.  In particular, two groups have recently reported on new
low-energy structure in the magneto-photoluminescence in GaAs structures
occurring at energies below the fundamental recombination from the
lowest Landau level (LLL).\cite{Gravier,Fink-prb97} While clearly
indicative of many-body effects, these new experimental results have led
to differing proposals for the mechanisms responsible for the various
spectral anomalies and a consensus has yet to be reached.  These
features have alternatively been attributed to a perturbative many-body
shake-up process\cite{Fink-prb97} and to a non-perturbative final state
resonance of the 2DEG.\cite{Gravier,Pawel4} It is the purpose of this
paper to present the results of a comprehensive experimental study of
the polarization, magnetic field and temperature dependence of these
optical anomalies and elucidate the distinct mechanisms responsible for
each.

We report on the observation of unusual low-energy structure in the
magneto-photoluminescence spectra for filling factors $\nu>2$.  We
identify and analyze two distinct features: (i) a doublet structure in
the fundamental recombination from the lowest Landau level develops at
odd integral filling factors $\nu=5$ and $\nu=3$, (ii) a significantly
smaller satellite peak which red shifts for increasing magnetic field
between $\nu=6$ and $\nu=2$ is observed at approximately 2meV below the
doublet structure.  While these features are separated by rather small
energy differences and occur roughly in the same regime of magnetic
field, we shall show that the data conclusively point to distinct
mechanisms for each process.

The two-dimensional electron system under investigation is
a single-side $n$-modulation doped Al$_x$Ga$_{1-x}$As-GaAs
250$\AA$ single quantum well (SQW).  Samples from three different wafers
have been studied with similar behavior seen in all samples.  All three
samples are of extremely high quality: electronic densities n$_s$ range
from $1.4-1.8 \cdot 10^{11} cm^{-2}$ and mobilities $\mu$ range from $
2.6-3.0\cdot 10^6 cm^2/Vs$.  The samples were excited at 740nm with the
output of a tunable Ti:sapphire laser.  The laser was stabilized
and the incident power density was kept below
$10^{-4}W/cm^{2}$.  We have verified that all measurements were made in
the linear regime by collecting data with incident power densities at
$10^{-6}W/cm^{2}$  with no change in the observed spectra.
Excitation delivery and signal collection were accomplished via an
optical fiber system that resides in He$^3$ refrigerator mounted in a
13T magnet.  Base temperature of our configuration is 500mK.  All
measurements were made in the transmission geometry and polarization
analysis is done {\sl in situ} with a circular polarizer placed
immediately following the sample but prior to the collection fiber.  The
signal is dispersed in a 0.64m monochromator and detected with a liquid
nitrogen-cooled charged-coupled-device (CCD) camera with 0.2meV
resolution.

Figure 1 displays the low temperature magneto-photoluminescence spectra
in the left-circular ($\sigma^-$) polarization in the field range of 0T
to 4T for a sample with n$_s$=$1.8 \cdot 10^{11}cm^{-2}$ and $\mu = 2.6
\cdot 10^6 cm^2/Vs$.  The use of polarization analysis is crucial since
the high quality of these samples results in narrow line-widths in which
spin-splittings are resolved above 1.5T.  Detecting in the $\sigma^-$
polarization guarantees that the fundamental recombination from the LLL
occurs between a $m_z=1/2$ electron level and a $m_z=3/2$ hole level.
The application of a quantizing magnetic field causes the zero field
spectrum to split into the Landau fan structure.  The gross behavior of
these spectra and its relationship to the quantum Hall effect have been
studied extensively.\cite{bbg-prl90} We focus our discussion on the low
energy structure between $\nu=5$ and $\nu=2$ appearing below the main
peak labeled LL$_0$ in Figure 2.  Within the context of a
non-interacting model, LL$_0$ is associated with the ground-state
recombination of an electron and hole, each sitting on the lowest Landau
level.  Figure 2 shows an individual spectra taken at $\nu=3$ and
T=0.53K in which two distinct low-energy features are visible below the
fundamental ground-state recombination LL$_0$.  Our analysis will focus
on the origin of these two optical anomalies, labeled OA$_0$ and OA$_1$.

We begin our discussion with the lowest energy feature, labeled OA$_1$,
which we attribute to the process of shake-up in a magnetic field.  In a
shake-up process a recombining electron-hole pair perturbs the electron
gas causing another electron to be excited across the cyclotron gap from
the lowest Landau level to the first Landau level.  Energy conservation
requires that the emitted photon's energy be lower than the fundamental
recombination by $\sim$ $\hbar \omega_c$.  Shake-up in a magnetic field
has been studied experimentally with similar behavior reported in
InGaAs-InP quantum wells\cite{Nash1}, and most recently, in GaAs
structures.\cite{Fink-prb97} The theory of shake-up has also been
extensively investigated.\cite{Pawel1,Pawel2} Most theoretical
approaches have assumed a significant amount of disorder in which
electron-electron interactions can be treated perturbatively and only
cyclotron gaps are considered.  Zeeman gaps, which are typically
unresolved in such systems, are ignored, and consequentially the
shake-up process is independent of the spin configuration of the 2DEG.
As shown in Figure 3, OA$_1$ moves to lower energy for increasing
magnetic fields with a slope approximately equal to -1.65meV/T,
corresponding to an inter-Landau level excitation in GaAs.\cite{mike} We
note that the absolute separation between OA$_1$ and LL$_0$ is actually
greater than $\hbar \omega_c$: at $\nu=3$ the recombinations are
separated by $\sim$ 6meV while $\hbar \omega_c$ at this field is
approximately 4meV.  This behavior has been observed previously in the
InGaAs-InP system\cite{Nash1} and attributed to the fact that for
non-zero wave-vectors the excitation energy of the n=1 magneto-plasmon
mode actually exceeds $\hbar \omega_c$.\cite{Kallin} For comparison, we
have drawn in Figure 3 a line through OA$_1$ that corresponds to an
excitation dispersing with a slope equal to the ideal (non-interacting)
Landau level separation in GaAs of 1.65meV/T.  The correspondence is
quite good.  The intensity of OA$_1$ is also quite small; its peak
intensity is only $\sim 6\%$ of the peak intensity seen in LL$_0$, and
between $\nu=6$ and $\nu=2$, OA$_1$ shows very little intensity
variation.  The smallness of the observed intensity in OA$_1$ and its
complete lack of significant variation over this field range clearly
identify it as a perturbative shake-up process.

We turn now to OA$_0$, the largest and most striking feature in Figure
2.  At $\nu=3$ OA$_0$ contains nearly 50 percent of the spectral weight
seen in the fundamental recombination band, LL$_0$.  The intensity of
OA$_0$ exhibits complicated magnetic field dependence which is clearly
seen in figure 4.  At $\nu=6$ there is no indication of the OA$_0$ but
at $\nu=5$ it is clearly developed.  This change happens quite
dramatically; the feature appears with a change of magnetic field of
only 0.1T.  The feature doesn't disappear at $\nu=4$ but rather its
intensity is suppressed relative to $\nu=5$ and it {\it blueshifts}
towards LL$_0$. OA$_0$ rapidly gains spectral weight at $\nu=3$, again
in a very narrow regime of magnetic field of approximately 0.1T.  For
higher magnetic fields the feature is greatly suppressed and largely
gone at $\nu=2$.  This behavior is summarized in Figure 3 where the
magnetic field dependence of all recombinations is displayed.  OA$_0$
does not move to lower energy for increasing field as one would
expect from a simple shake-up process: it shows very little dispersion
except for blueshifts seen at $\nu=4$ and $\nu=2$.  The temperature
dependence of this novel structure is also quite telling and is
displayed in Figure 5.  OA$_0$, which appears as broad shoulder to the
LL$_0$ recombination at T=4.2K, doesn't change significantly down to
temperatures of 2K.  The most dramatic changes occur between 1.5K and
0.53K.  This energy scale of 1K $\sim$ 0.08meV is much smaller than any
cyclotron gap and indicates the importance of many-body and/or spin
effects.

The most striking aspects of OA$_0$ are its huge intensity at odd
filling factor $\nu=3$ and the complex dependence of its intensity and
energy position on magnetic field.  The strong resonance seen at odd
filling suggests that the spin of the electron system plays an important
role.  Our understanding of this doublet structure follows the recently
developed model of Hawrylak {\it et al.}\/\cite{Pawel4} which has been
used to explain a similar behavior seen in the data of Gravier and
co-workers.\cite{Gravier} This model has the advantage that it is
applicable in the limit of low disorder in which both Zeeman and
cyclotron gaps are resolved and electron-electron interactions are
expected to be of prime importance.  The recombination of a
conduction-band electron and valence-band hole leaves a hole in the
final state of the electron system.  In the non-interacting limit this
hole represents a well-defined quasiparticle.  Nevertheless, this hole
left behind in the final state of the 2DEG lies well below the Fermi
level and constitutes a highly excited state.  If this hole is
degenerate with other elementary excitations of the electron gas and
electron-electron interactions are strong, the spectral function of the
fully interacting hole may not be perturbatively related to the hole in
the non-interacting system.  This is the fundamental finding of
Hawrylak's work: the doublet structure observed in luminescence at odd
filling factors is due to the non-quasiparticle behavior of the 2DEG
hole spectral function.  The two peaks observed in luminescence are
shown to be due to a resonant many-body interaction of the hole in the
electron gas with a continuum of spin-wave excitations.  Thus spin plays
a crucial role in the physics of this model; the splitting of the
spectral function only occurs at odd fillings where a continuum of
low-lying spin waves is resonant and the spin of the hole can be
compensated for by a spin-flip excitation.  Hawrylak has deemed this
situation an ``Anderson-Fano'' resonance in which the hole strongly
interacting with the continuum of low-lying spin waves is mapped onto
the classic solid-state physics problem of a localized state interacting
with an unbound continuum.

We believe that such theoretical considerations describe qualitatively
our experimental observations of the doublet LL$_0$ and OA$_0$, and
clearly distinguish it from the perturbative shake-up process OA$_1$.
The hole spectral function has been calculated numerically\cite{Pawel4}
and the lower energy peak of the doublet is found to contain 60 percent
of the spectral weight of the fundamental recombination band at $\nu=3$.
This splitting into a resonant doublet structure is consistent with the
experimentally observed behavior.  Additionally this theory accounts
well for the observed blueshifts and intensity reductions seen in OA$_0$
at even filling $\nu=4$ and $\nu=2$ where the spectral function is
expected to collapse to a single peak in the absence of low-lying
intra-Landau level spin wave excitations.  Both experiment and theory
point to the important role played by the spin magnetization of 2DEG for
final state interactions.

In summary, we have presented a systematic experimental study of the low
energy structure of the photoluminescence spectra from a low disorder
2DEG at integral filling $\nu>2$.  Our findings identify two distinct
features occurring below the fundamental band of recombination from the
lowest Landau level.  The lowest energy satellite is consistent with a
perturbative shake-up process in a magnetic field.  The splitting of the
fundamental recombination line at odd filling factors is shown to be
related to a non-perturbative splitting of the final state spectral
function of the 2DEG.

\section*{Acknowledgments}
This work was supported by National Science Foundation Grant No.
DMR-9701958.  We thank Pawel Hawrylak for many insightful conversations.

\newpage

\begin{center}
{\Large FIGURES}
\end{center}

\vspace{0.5in}

\noindent
{\bf Figure 1.} Photoluminescence spectra in the LCP ($\sigma^-$)
polarization at T=0.53K for the SQW with $ n_s=1.8\cdot 10^{11} cm^{-2}$
and $\mu = 2.6 \cdot 10^6 cm^2/Vs$.  The spectra at $\nu=3$ is
highlighted to show the low-energy fine-structure.  The inset to Fig. 1
displays the lowest-energy feature tuning to lower energy for increasing
magnetic field.

\vspace{0.5in}

\noindent
{\bf Figure 2.}  Individual spectra at $\nu=3$ displaying low energy
anomalies OA$_0$ and OA$_1$.  The fundamental recombinations from the
0th and 1st Landau level are labeled LL$_0$ and LL$_1$ respectively.
Note the strong intensity of the OA$_0$ peak relative to the fundamental
LL$_0$. 

\vspace{0.5in}

\noindent
{\bf Figure 3.}  Recombination peak positions as function of magnetic
field at T=0.53K. The low energy features are labeled OA$_0$ and OA$_1$.
The fundamental recombination from the 0th Landau level is labeled
LL$_0$.  Also shown is the line E=-1.65meV/T $\cdot$B+E$_0$ as explained
in the text.

\vspace{0.5in}

\noindent
{\bf Figure 4.} Magnetic field development of optical anomalies between
$\nu=6$ and $\nu=2$.  All spectra at taken in the $\sigma^-$
polarization at T=0.53K.  For ease of comparison, all spectra have been
normalized to have the same peak intensity in the LL$_0$ transition.

\vspace{0.5in}

\noindent
{\bf Figure 5.} Temperature dependence of low energy structure between
4.2K and 0.53K.  All scans are taken at B=2.5T, $\nu=3$.

\newpage

\newbox\figa
\setbox\figa=\vtop{\kern0pt\psfig{figure=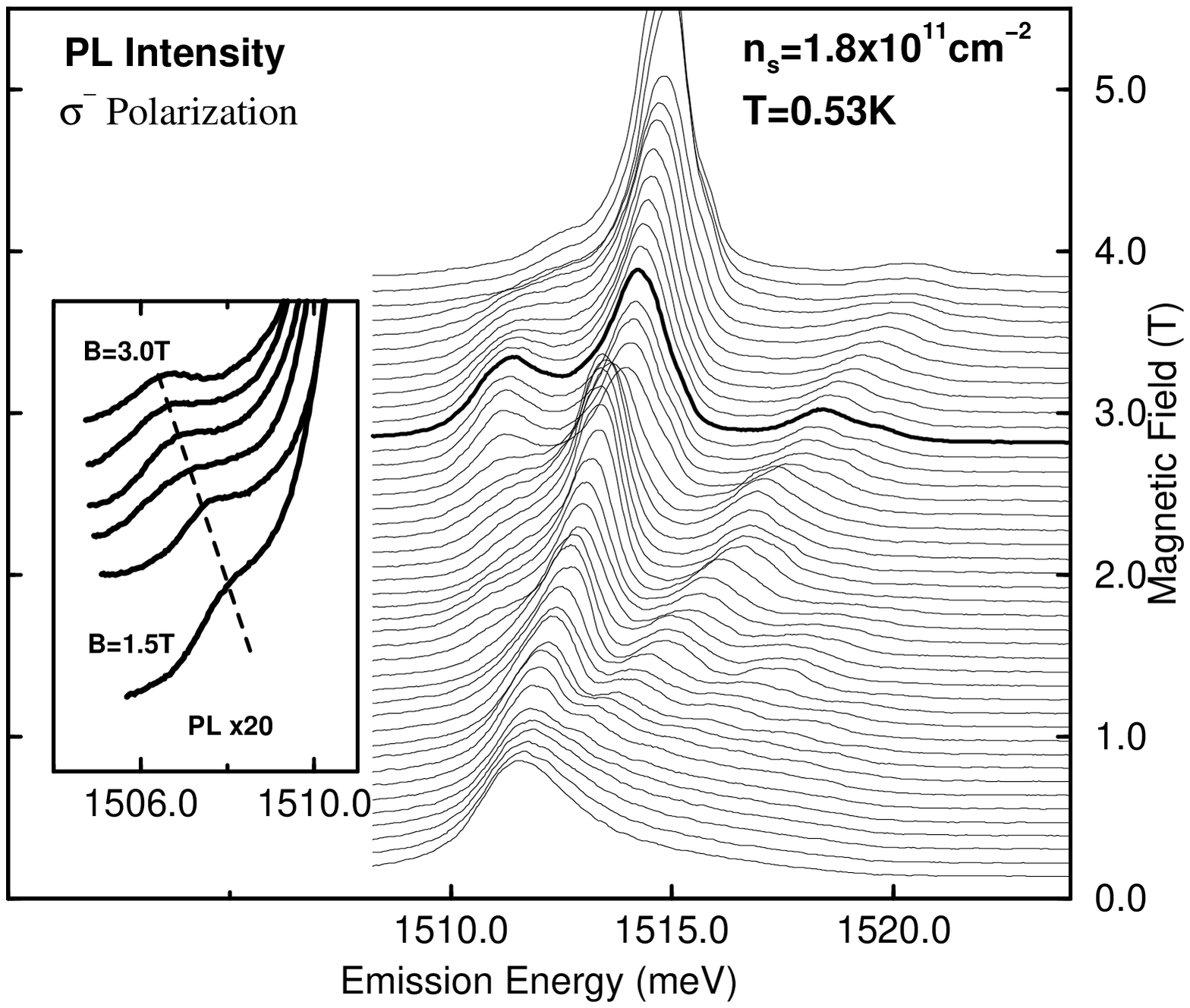,width=6.0in}}

\centerline{\box\figa}

\vfill
\centerline{Manfra, {\sl et al.} Fig 1.}

\newpage

\newbox\figb
\setbox\figb=\vtop{\kern0pt\psfig{figure=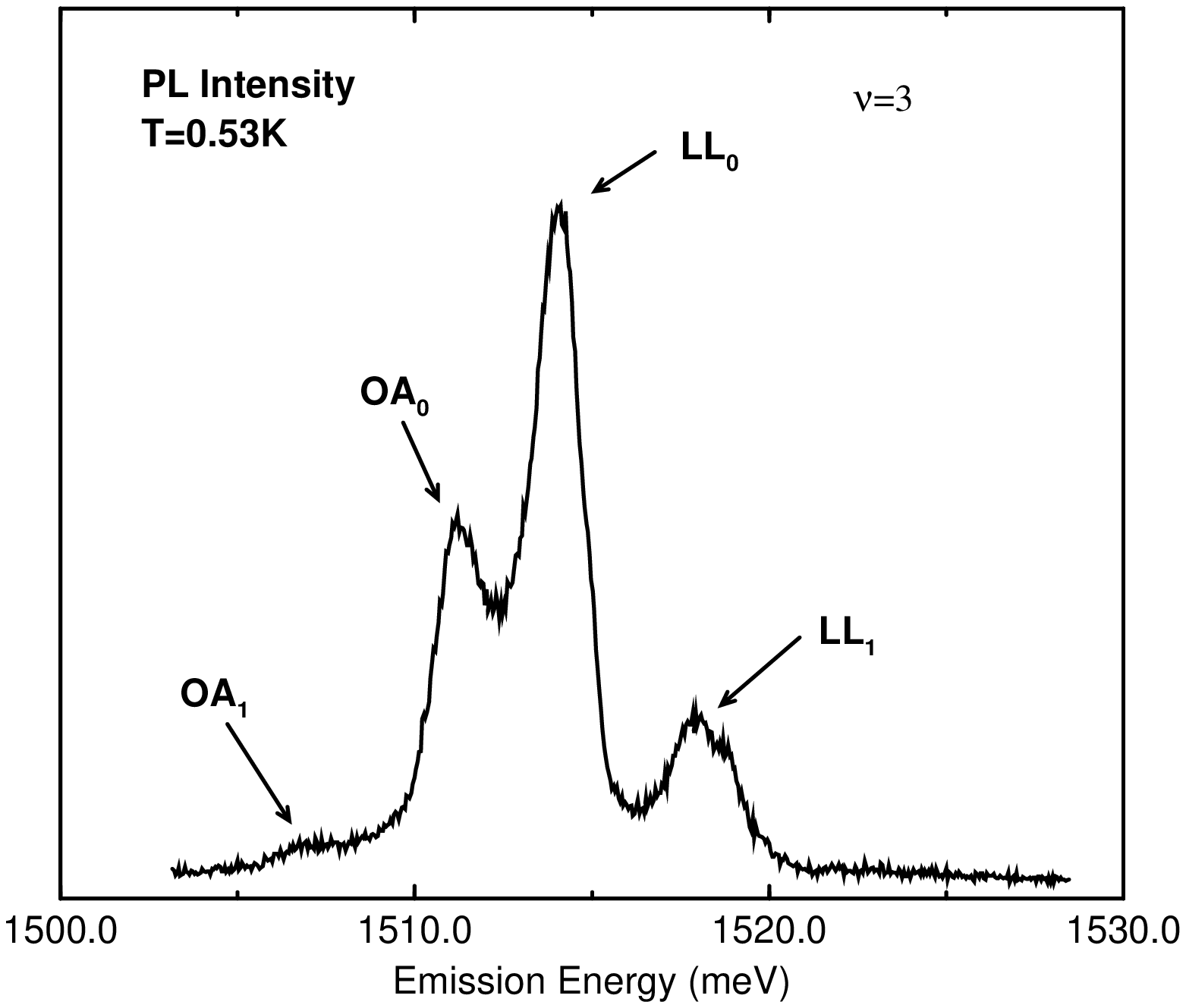,width=6.0in}}

\centerline{\box\figb}
\vfill
\centerline{Manfra, {\sl et al.} Fig 2.}
\newpage

\newbox\figc
\setbox\figc=\vtop{\kern0pt\psfig{figure=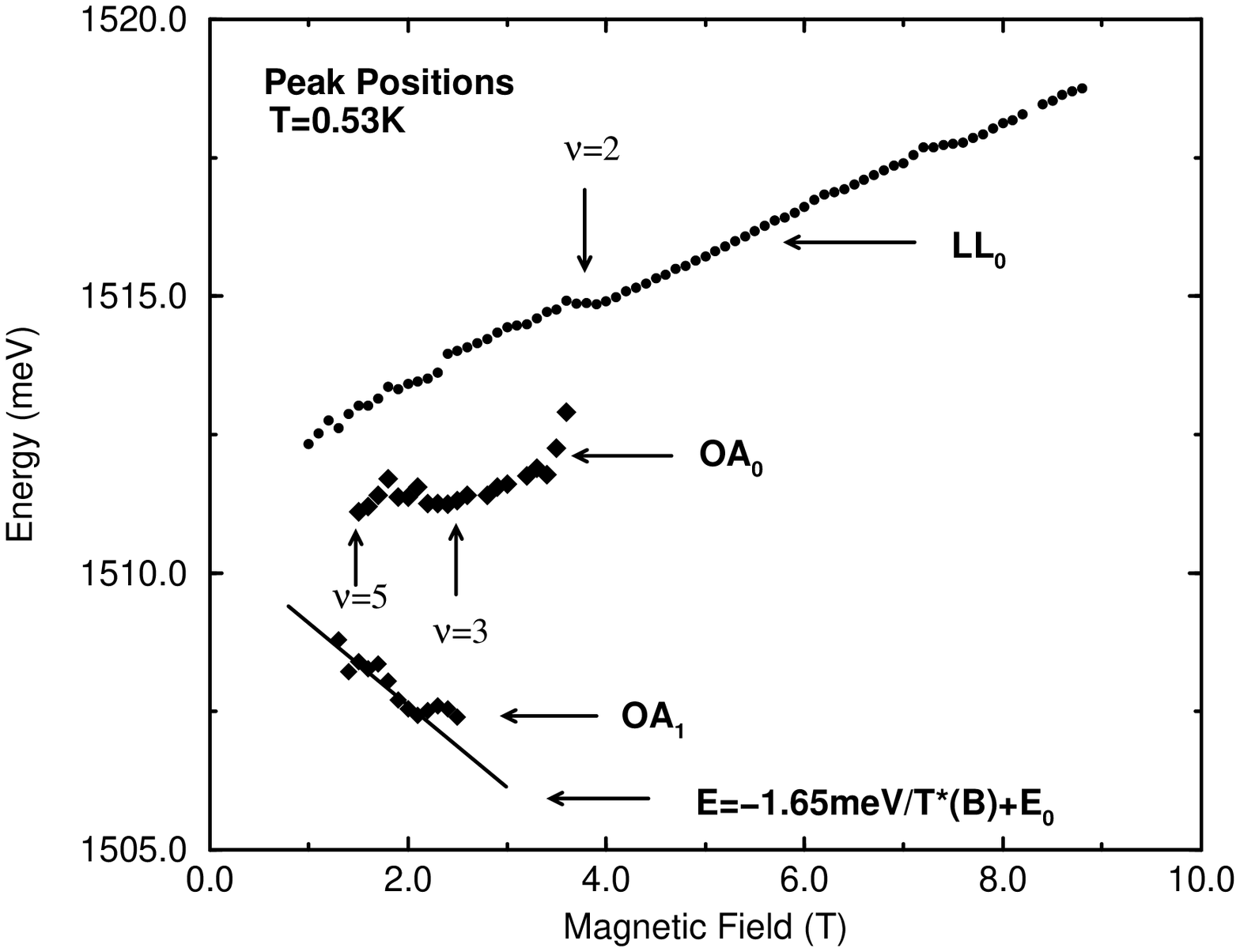,width=6.0in}}

\centerline{\box\figc}
\vfill
\centerline{Manfra, {\sl et al.} Fig 3. }

\newpage

\newbox\figd
\setbox\figd=\vtop{\kern0pt\psfig{figure=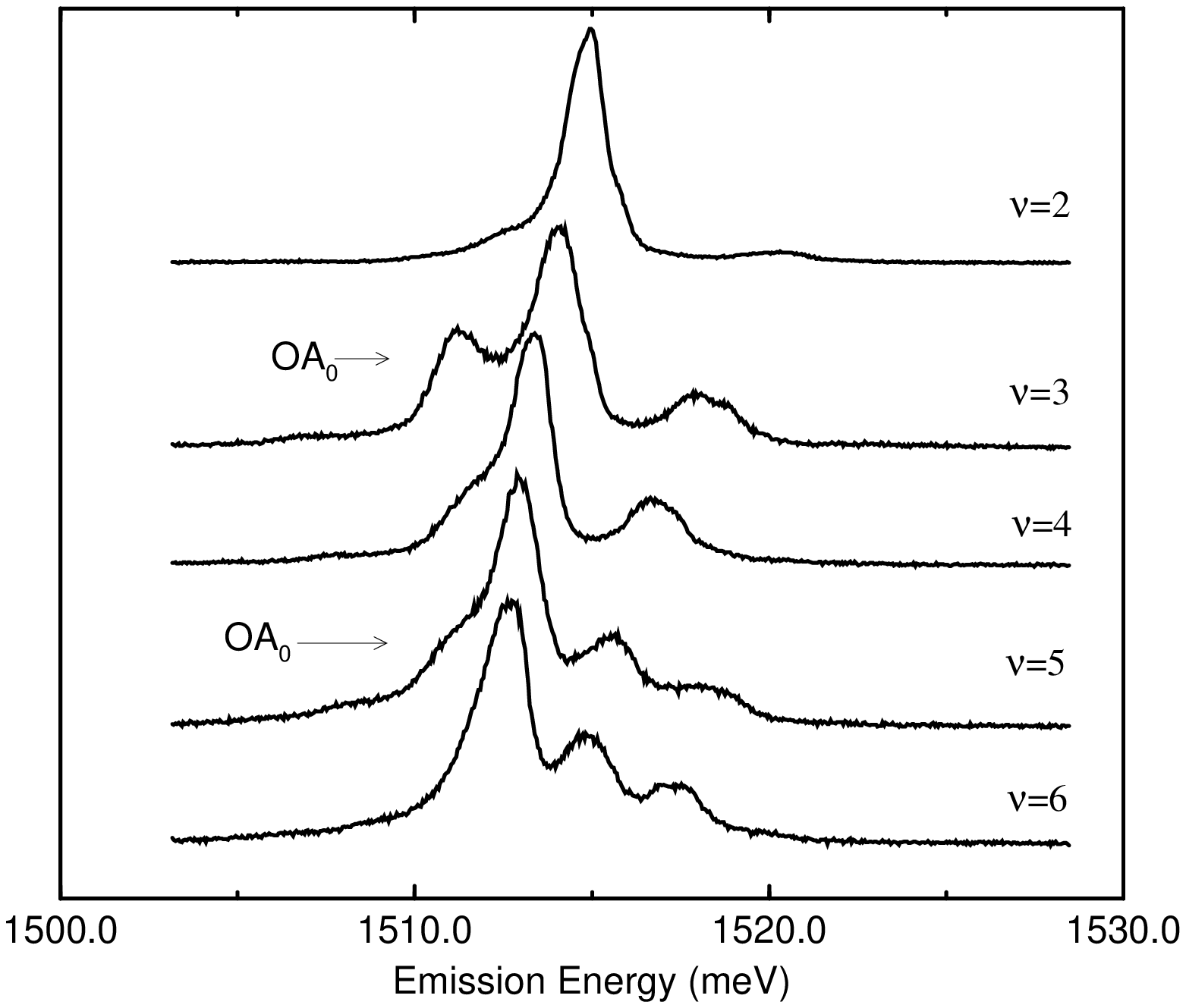,width=6.0in}}

\centerline{\box\figd}
\vfill
\centerline{Manfra, {\sl et al.} Fig 4.}
 
\newpage

\newbox\fige
\setbox\fige=\vtop{\kern0pt\psfig{figure=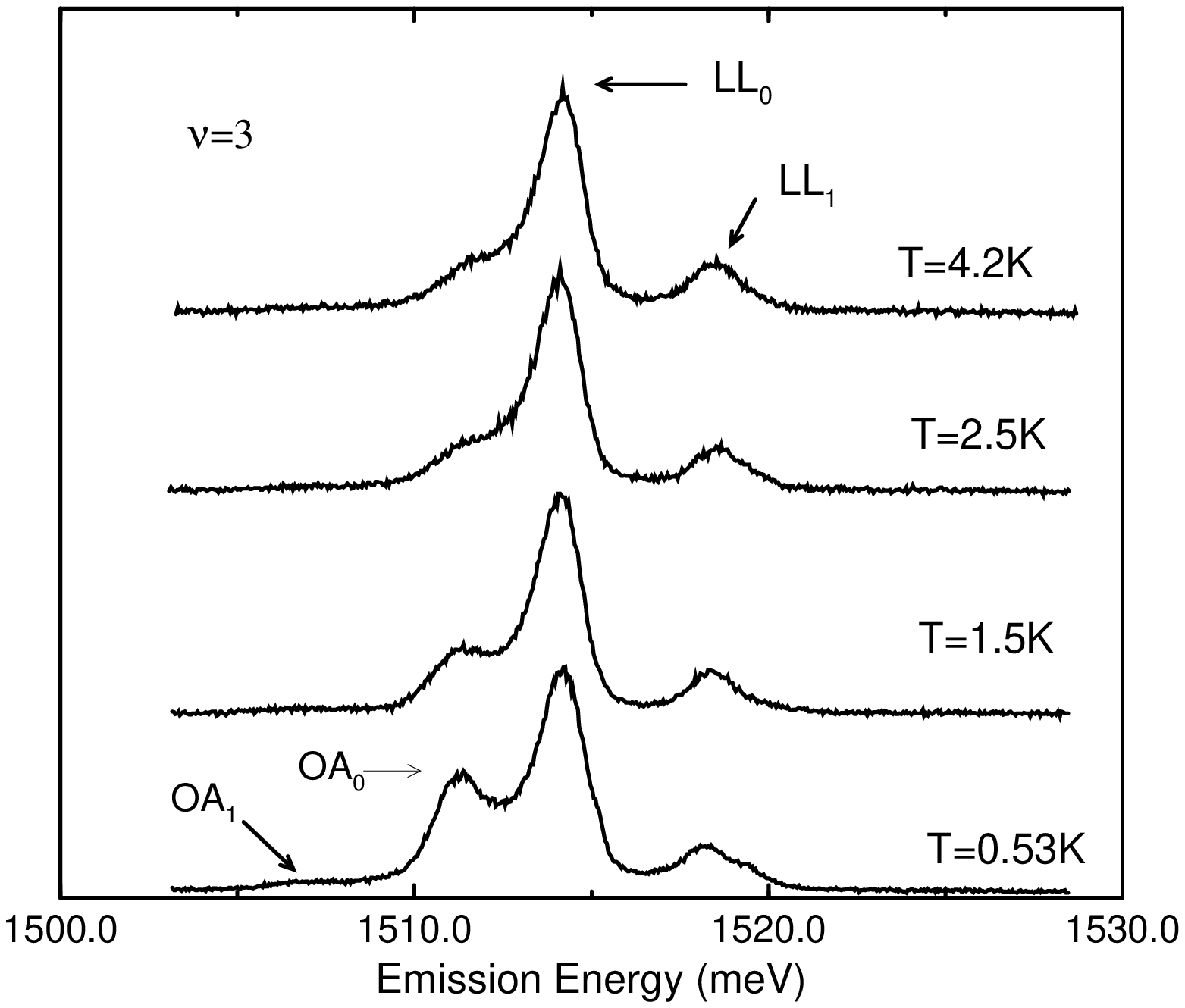,width=6.0in}}

\centerline{\box\fige}
\vfill
\centerline{Manfra, {\sl et al.} Fig 5.}


\begin{thebibliography}{99}
\bibitem{mike}
Using an effective mass m$^*$ = 0.0699m$_e$ for GaAs, the separation
between Landau levels is $\hbar \omega_c $ =
1.65meV/T $\cdot$ B.
\end{thebibliography}
\end{document}